\documentclass[12pt]{article}

\begin{document}

\input amssym.tex

\title{Polarized vector bosons on the de Sitter expanding universe}

\author{Ion I. Cot\u aescu  \\
{\small \it West University of Timi\c soara,}\\
{\small \it V. Parvan Ave. 4 RO-300223 Timi\c soara,  Romania}}

\maketitle

\begin{abstract}
The quantum theory of the vector field minimally coupled to the gravity of the
de Sitter spacetime is built in a canonical manner starting with a new complete
set of quantum modes of given momentum and helicity derived in the moving chart
of conformal time. It is shown that the canonical quantization leads to new
vector propagators which satisfy similar equations as the propagators derived
by Tsamis and Woodard [{\em J.Math.Phys.} {\bf 48} (2007) 052306] but having a
different structure. The one-particle operators are also written down pointing
out that their properties are similar with those found already in the quantum
theory of the scalar, Dirac and Maxwell free fields.

 Pacs: 04.62.+v
\end{abstract}

Keywords: vector field; de Sitter spacetime; canonical quantization;
propagators; one-particle operators.

%\maketitle

\newpage

\section{Introduction}

In the quantum theory of fields on curved spacetimes the de Sitter (dS)
expanding universe carrying fields variously coupled to gravity is of a special
interest \cite{BD,WALD}. The free scalar and Dirac fields minimally coupled to
gravity were studied in static charts as well as in the (co)moving charts with
proper or conformal times. The free massive vector field is considered only in
static charts \cite{V} with spherical coordinates and, therefore, the problem
of its quantum modes in moving charts is still open.

In the simplest case of scalar fields the quantum theory leads to propagators
which depend on the geodesic length \cite{PPP,PPG}. This fact suggests that the
difficulties arising in the theory of vector fields can be avoided in an
elegant manner considering directly two-point functions depending only on the
geodesic length and its derivatives \cite{PPV,PPV1}. By using this method one
skips the principal steps of the traditional quantum theory based on the
canonical quantization and, consequently,  some delicate problems may remain
unsolved. An example is the uniqueness of the spin for which we have not yet a
coherent theory in general relativity. In other respects, the flat limit of the
quantum field theory on curved spacetimes must recover the traditional theory
where the canonical quantization is working. For this reason we emphasize  that
it deserves to construct the canonical quantum field theory on dS manifolds
following the same steps as in special relativity. In this approach, apart from
propagators, one may derive the conserved one-particle operators which are
crucial for the physical interpretation of the field quanta (determining the
charge, spin, etc.).

To this aim we have then all the elements we need. First of all we can
analytically solve the equations of the principal fields minimally coupled to
the dS gravity. Moreover, the $SO(1,4)$ symmetry of the dS manifolds provides
us with a large collection of operators commuting with those of the field
equations \cite{CML,ES}. Thus we can choose complete sets of commuting
operators which should determine suitable systems of fundamental solutions as
common eigenfunctions of the operators of these sets. Normalizing  the
solutions with respect to a well-defined relativistic scalar product, we may
then obtain the complete systems of fundamental solutions we need for
canonically quantizing the free fields. On the dS spacetime the momentum
operators commute with that of the field equation and, therefore, there are
fundamental solutions representing eigenfunctions of the momentum operators.
Such solutions form the momentum basis in which the Hamiltonian operator is not
diagonal since it does not commute with the momentum operators. In this
context, we proposed a new time-evolution picture \cite{CSP} which allowed us
to introduce the energy basis which completes  the quantum theory of the scalar
field \cite{CS} on the dS expanding universe. Within the same conjecture we
developed the quantum theory of the Dirac \cite{CD1,CD2} and Maxwell \cite{Max}
fields.  What then remains is to study the Proca theory of the vector fields
whose propagators  or two-point functions in moving charts are of actual
interest in cosmology \cite{Dav,Woo}. For this reason, the canonical quantum
theory of the vector field on dS moving charts represents the subject of the
present paper.

We start in the second section with a brief review of the Proca theory  on the
moving dS charts with conformal time, by introducing then the principal
conserved operators generated by the Killing vectors. Furthermore, in section 3
we derive the complete set of fundamental solutions of the field equation
determined by momentum and helicity. The next section is devoted to the
canonical quantization of the vector field which leads to the new Green
functions we look for. In section 5 we derive the principal one-particle
operators in momentum representation.

The principal results of this paper are the vector quantum modes of given
momentum and helicity, the one-particle operators and the vector propagators of
the canonical quantum theory on the dS spacetime. We must stress that these
propagators are different from the maximally symmetric two-point functions
derived before  by Allen and Jacobson \cite{PPV} but satisfy the equations
proposed by  Tsamis and Woodard \cite{PPV1}. The arguments we present here
indicate that our propagators are new solutions of these equations.

\section{Preliminaries}

In a given  local chart of coordinates $x^{\mu}$ ($\mu,\nu,...=0,1,2,3 $) and
the line element
\begin{equation}
ds^{2}=g_{\mu \nu}d{x}^{\mu}d{x}^{\nu}\,,
\end{equation}
of a curved spacetime, $(M,g)$, the Proca theory of the massive charged vector
field $A$ minimally coupled to gravity has the action
\begin{equation}\label{action}
\mathcal{S}[A]=\int d^{4}x \sqrt{g}\,{\cal L}=\int d^{4}x
\sqrt{g}\,\left[-\frac{1}{2}\,F_{\mu \nu  }^*F^{\mu \nu  } +m^2
A_{\mu}^*A^{\mu}\right],
\end{equation}
where $m$ is the mass, $g=|\det(g_{\mu\nu})|$ and $F_{\mu \nu
}=\partial_{\mu } A_{\nu }-\partial_{\nu } A_{\mu }$ is the field
strength. From the resulted field equations,
\begin{equation}\label{EQ}
\partial_{\nu}(\sqrt{g}\,g^{\nu\alpha}g^{\mu\beta}F_{\alpha\beta})
+m^2 \sqrt{g}\, A^{\mu}=0\,,
\end{equation}
one deduces the Lorentz (or transversality) condition
\begin{equation}
\partial_{\mu}(\sqrt{g} A^{\mu})=0\,,
\end{equation}
which guarantees the uniqueness of the spin $s=1$.

The whole theory is invariant under the $U(1)$ internal symmetry
transformations (generated by the identity operator $I$) and the isometries
related to the Killing vectors of $(M,g)$. For each isometry transformation
$x\to x'=\phi_{\xi}(x)$ depending on the group parameter $\xi$ there exists an
associated Killing vector field, ${K}=\partial_{\xi}\phi_{\xi}|_{\xi=0}$ (which
satisfy the Killing equation ${K}_{\mu;\nu}+{K}_{\nu;\mu}=0$). Under such
isometry the vector field transforms as $A\to A' =T_{\xi}A$, according to the
operator-valued representation $\xi\to T_{\xi}$ of the isometry group defined
by the well-known rule
\begin{equation}
\frac{\partial \phi^{\nu}_{\xi}(x)}{\partial
x_{\mu}}\left(T_{\xi}A\right)_{\nu}[\phi(x)]=A_{\mu}(x)\,.
\end{equation}
The corresponding generator, $X_{K}=i\,\partial_{\xi}T_{\xi}|_{\xi=0}$, has the
action
\begin{equation}\label{XA}
(X_{K}\, A)_{\mu}=-i({K}^{\nu}A_{\mu;\nu}+{K}^{\nu}_{~;\mu}A_{\nu})\,.
\end{equation}
We say that these generators are conserved operators since they
commute with the operator of the field equation \cite{ES}. Moreover,
from the Noether theorem it results that any symmetry generator $X$
gives rise to the time-independent quantity,
\begin{equation}\label{Ck}
C[{X}]=-{i}\int_{\Sigma}d\sigma^{\mu}\sqrt{g}\,g^{\alpha\beta}\left[
A_{\alpha}\stackrel{\leftrightarrow}{\partial_{\mu}} \,(X
A_{\beta})\right]\,,
\end{equation}
on a given space-like hypersurface $\Sigma \subset M$. Particularly,
for the internal $U(1)$ symmetry we must take  $X=I$.

We consider $(M,g)$ to be the dS spacetime defined as a hyperboloid of radius
$1/\omega$ \footnote{We denote by $\omega $ the Hubble dS constant since  $H$
is reserved for the Hamiltonian operator} in a five-dimensional
pseudo-Euclidean manifold, $M^5$, of coordinates $z^A$ labeled by the indices
$A,\,B,...= 0,1,2,3,5$. The local charts of coordinates $\{x\}$ on $M^5$ can be
easily introduced giving the specific functions $z^A(x)$. Here we consider only
the moving chart $\{t,{\bf x}\}$ with the {conformal} time, $t$, Cartesian
coordinates  and  the line element
\begin{equation}
ds^{2}=\frac{1}{(\omega t)^2}\eta_{\mu \nu} dx^{\mu}
dx^{\nu}=\frac{1}{(\omega t)^2}\,\left(dt^{2}- d{\bf x}\cdot d{\bf
x}\right)\,,
\end{equation}
where $\eta$ is the the metric tensor of the Minkowski spacetime
\cite{BD}.

The isometry group of the dS manifold is just the group $SO(1,4)$ of the
pseudo-orthogonal transformations in $M^5$. For this reason, the
basis-generators of the $SO(1,4)$ algebra are associated to ten independent
Killing vectors, ${K}_{(AB)}=-{K}_{(BA)}$, which give rise to the
basis-generators $X_{(AB)}$ of the vector representation of the $SO(1,4)$ group
carried by the space of the vector fields $A$. In what follows  we focus only
on the Hamiltonian (or energy) operator $H=\omega X_{(05)}$, the momentum
components $P^i=\omega(X_{(5i)}-X_{(0i)})$ and those of the total angular
momentum $J_i=\frac{1}{2}\varepsilon_{ijk}X_{(jk)}$ ($i,j,...=1,2,3$)
\cite{CD1}. The action of these operators can be deduced from Eq. (\ref{XA})
using the concrete form of the corresponding Killing vectors whose components
read \cite{ES}: $ K^{\mu}_{(05)}=x^{\mu}$ and
\begin{eqnarray}
K^0_{(5i)}-K^0_{(0i)}=0\,,& \quad &K^j_{(5i)}-K^j_{(0i)}=
\frac{1}{\omega}\,\delta_{ij}\,,\\
K^0_{(ij)}=0\,,& &K^k_{(ij)}=\delta_{ki}x^j-\delta_{kj}x^i\,.
\end{eqnarray}
The Hamiltonian and momentum operators do not have spin parts,
acting as
\begin{eqnarray}
(H\, A)_{\mu}(t,\bf{x})&=&-i\omega(t \partial_t + x^i\partial_i +1
) {A}_{\mu}(t,\bf{x})\,,\label{HH}\\
(P^i A)_{\mu}(t,\bf{x})&=&-i\partial_i\,A_{\mu}(t,\bf{x})\,,\label{PP}
\end{eqnarray}
while the action of the total angular momentum reads
\begin{eqnarray}
(J_i\, A)_j(t,\bf{x})&=&(L_i A)_j(t,{\bf x})
-i\varepsilon_{ijk}A_k(t,{\bf x})\,,\label{J}\\
(J_i\, A)_0(t,\bf{x})&=&(L_i A)_0(t,{\bf x})\,,
\end{eqnarray}
where ${\bf L}={\bf x}\times {\bf P}$ is the usual angular momentum operator.
In addition, we define the Pauli-Lubanski operator $W={\bf P}\cdot {\bf J}$
whose action depends only on the spin parts,
\begin{equation}
(W A)_i(t,{\bf x})=\varepsilon_{ijk}\partial_j A_k(t,{\bf x})\,, \quad (W
A)_0(t,{\bf x})=0\,.
\end{equation}
This operator will define the  polarization in the canonical basis of the
$so(3)$ algebra as in special relativity.

Starting with the above results we can derive the time-independent quantities
defined by Eq. (\ref{Ck}) for any symmetry generator $X$ and $\Sigma = {\Bbb
R}^3$. After a little calculation we obtain the compact form
\begin{equation}\label{Ck1}
C[X]=-\eta^{\mu\nu}\, \left<A_{\mu},(XA)_{\nu}\right>\,,
\end{equation}
with the new notation
\begin{equation}\label{fg}
\left<f,g\right>=i\int d^3x\, f^*(t,{\bf x})
\stackrel{\leftrightarrow\,\,\,}{\partial_{t}} g(t,{\bf x})\,,
\end{equation}
where $f\stackrel{\leftrightarrow}{\partial}g=f(\partial g)-g(\partial f)$.
Defining now the relativistic scalar product of two vector fields as \cite{BD}
\begin{equation}\label{AA}
\left<A|\,A'\right>=-\eta^{\mu\nu}\left<A_{\mu},A'_{\nu}\right>=-i\eta^{\mu\nu}\int
d^3x\, A^*_{\mu}(t,{\bf x})
\stackrel{\leftrightarrow\,\,\,}{\partial_{t}} A'_{\nu}(t,{\bf x})\,
 \,,
\end{equation}
we can write
\begin{equation}\label{CX}
C[X]=\left<A|\,XA\right>.
\end{equation}

\section{Polarized plane wave solutions}

The specific feature of the quantum mechanics on dS manifolds is that the
Hamiltonian operator (\ref{HH}) does not commute with the momentum operators
(\ref{PP}). For this reason the particular solutions of the field equation may
be eigenfunctions either of the momentum operators or of the Hamiltonian one.
In what follows  we derive a complete set of fundamental solutions as common
eigenfunctions of the commuting operators $P^i$ and $W$.

In the chart $\{t,{\bf x}\}$ the field equations take the form
\begin{eqnarray}
&&\partial_t(\partial_i A_i)-\Delta
A_0+\frac{\mu^2}{t^2}\,A_0=0 \,,\label{EdS1}\\
&&\partial_t^2A_k-\Delta A_k-\partial_k(\partial_t A_0)+\partial_k(\partial_i
A_i)+\frac{\mu^2}{t^2}\,A_k=0\,,\label{EdS2}
\end{eqnarray}
where $\mu=m/\omega$, while the Lorentz condition reads
\begin{equation}\label{Lor}
\partial_i A_i =\partial_t A_0-\frac{2}{t}\,A_0\,.
\end{equation}
The solutions of these equations are vector fields which can be expanded as,
\begin{eqnarray}
A&=&A^{(+)}+A^{(-)}\nonumber\\
&=&\int d^3p \sum_{\lambda}\left\{{\rm U}[{\bf p},\lambda] a({\bf p},\lambda)+
{\rm U}[{\bf p},\lambda]^* b^*({\bf p},\lambda)\right\}\,,\label{field1}
\end{eqnarray}
in terms of wave functions in momentum representation, $a({\bf p},\lambda)$,
and $b({\bf p},\lambda)$, which depend on the momentum ${\bf p}\in {\Bbb
R}^3_p$ and the polarization $\lambda=0,\pm 1$. Denoting $p=|{\bf p}|$ we
assume that the vector fields ${\rm U}[{\bf p},\lambda]$ satisfy the eigenvalue
equations
\begin{equation}
P^i {\rm U}[{\bf p},\lambda]=p_i {\rm U}[{\bf p},\lambda]\,, \quad W {\rm
U}[{\bf p},\lambda]=p\,\lambda {\rm U}[{\bf p},\lambda]\,,
\end{equation}
and the orthonormalization relations
\begin{equation}\label{orto}
\left<{\rm U}[{\bf p},\lambda]|\,{\rm U}[{\bf
p}',\lambda']\right>=\delta_{\lambda\lambda'}\delta^3({\bf p}-{\bf p}')\,.
\end{equation}

The components of these vector fields, U$[{\bf p},\lambda]_{\mu}(x)$, form the
desired system of fundamental plane wave solutions of positive frequencies
depending on momentum and polarization. The corresponding fundamental solutions
of negative frequencies are  U$[{\bf p},\lambda]_{\mu}(x)^*$. We suppose that
these solutions are of the form
\begin{equation}\label{UU1}
{\rm U}[{\bf p},\lambda]_{i}(x)=\left\{
\begin{array}{lcl}
\alpha(t,p)\, e_i({\bf n}_p,\lambda)\, e^{i{\bf p}\cdot{\bf x}}&{\rm for}&\lambda=\pm 1\\
\beta(t,p)\, e_i ({\bf n}_p,\lambda)\,e^{i{\bf p}\cdot{\bf x}}&{\rm
for}&\lambda=0
\end{array}\right.
\end{equation}
and
\begin{equation}\label{UU2}
{\rm U}[{\bf p},\lambda]_{0}(x)=\left\{
\begin{array}{lcl}
0&{\rm for}&\lambda=\pm 1\\
\gamma(t,p)\,e^{i{\bf p}\cdot{\bf x}}&{\rm for}&\lambda=0
\end{array}\right.
\end{equation}
where ${\bf n}_p={\bf p}/p$ and $e_i({\bf n}_p,\lambda)$ are the polarization
vectors of the helicity basis. For $\lambda=0$ these vectors are longitudinal,
i.e. ${\bf e}({\bf n}_p,0)={\bf n}_p$, while the vectors with $\lambda=\pm 1$
are transversal such that ${\bf p}\cdot{\bf e}({\bf n}_p,\pm 1)=0$. In general,
they have  c-number components which must satisfy \cite{BS,SW1}
\begin{eqnarray}
{\bf e}({\bf n}_p,\lambda)^*\cdot{\bf e}({\bf
n}_p,\lambda')&=&\delta_{\lambda\lambda'}\,,\\
{\bf e}({\bf n}_p,\lambda)^*\land {\bf e}({\bf
n}_p,\lambda)&=&i \lambda\, {\bf n}_p\,,\label{eee}\\
 \sum_{\lambda}e_i({\bf n}_p,\lambda)^*\,e_j({\bf
n}_p,\lambda)&=&\delta_{ij}\,.\label{tran}
\end{eqnarray}

Introducing now the functions (\ref{UU1}) and (\ref{UU2}) in Eqs. (\ref{EdS1}),
(\ref{EdS2}) and (\ref{Lor}) we obtain the system
\begin{eqnarray}
&&\left(\frac{d^2}{dt^2}+p^2+\frac{\mu^2}{t^2}\right)\alpha(t,p)=0\label{eqa}\\
&&\left(\frac{d^2}{dt^2}
-\frac{2}{t}\frac{d}{dt}+p^2+\frac{\mu^2+2}{t^2}\right)\gamma(t,p)=0\label{eqb}\\
&&\beta(t,p)=-\frac{i}{p}\left(\frac{d}{dt}-\frac{2}{t}\right)\gamma(t,p)\label{eqc}
\end{eqnarray}
which can be solved in terms of Bessel functions. The solutions read
\begin{eqnarray}
\alpha(t,p)&=&N_1 e^{-\frac{1}{2}\pi k}(-t)^{\frac{1}{2}}\,H^{(1)}_{ik}(-pt)\,,\label{alpha}\\
\gamma(t,p)&=&N_2 e^{-\frac{1}{2}\pi k}(-t)^{\frac{3}{2}}\,H^{(1)}_{ik}(-pt)\,,\label{gamma}\\
\beta(t,p)&=&iN_2e^{-\frac{1}{2}\pi
k}\left[\frac{1}{p}\left(ik+\frac{1}{2}\right)(-t)^{\frac{1}{2}}H^{(1)}_{ik}(-pt)
\right.\nonumber\\
&&\hspace*{36mm}\left.-(-t)^{\frac{3}{2}}H^{(1)}_{ik+1}(-pt)\right]\,,
\end{eqnarray}
where $H^{(1)}$ are Hankel functions, $N_1$ and $N_2$ are normalization factors
and  $k=\sqrt{\mu^2-\frac{1}{4}}$ provided $m>\omega/2$. By using then the
formulas given in the Appendix A, we find that the orthonormalization condition
(\ref{orto}) is accomplished only if we take (up to phase factors)
\begin{equation}\label{N1N2}
N_1=\frac{\sqrt{\pi}}{2}\frac{1}{(2\pi)^{3/2}}\,,\quad
N_2=\frac{\sqrt{\pi}}{2}\frac{1}{(2\pi)^{3/2}}\frac{\omega p}{m}\,.
\end{equation}
With this normalization our solutions satisfy the identity
\begin{eqnarray}
&&i \sum_{\lambda}\int d^3p\,{\rm U}[{\bf p},\lambda]_{i}^*(t,{\bf
x})\stackrel{\leftrightarrow\,\,\,}{\partial_{t}} {\rm U}[{\bf
p},\lambda]_{j}(t,{\bf x}')\nonumber\\
&&\hspace*{38mm}= \left(-\delta_{ij}+\frac{\omega^2
t^2}{m^2}\,\partial_i\partial_j\right)\delta^3({\bf x}-{\bf x}')\,,\label{comp}
\end{eqnarray}
which plays the role of a completeness condition. A similar equation can be
derived for the components $(0,0)$ but for $(0,i)$ and $(i,0)$ there are
imaginary parts that can not be evaluated. However, this is not an impediment
since the Lorentz condition reduces the number of canonical variable and,
therefore, Eq. (\ref{comp}) is enough for testing the completeness.

We derived thus the complete system of orthonormalized fundamental solutions
which are common eigenfunctions of the complete set of commuting operators
$\{P^i,W\}$. The solution of positive and negative frequencies correspond to
the eigenvalues $\{p^i,p\lambda\}$ and respectively  $\{-p^i,-p\lambda\}$. We
note that the massless limit makes sense only if we take
$\beta(t,p)=\gamma(t,p)=0$ which leads to the Coulomb gauge of the Maxwell free
field \cite{Max}.

\section{Quantization and propagators}

The quantization can be done in canonical manner transforming the wave
functions $a$ and $b$ of the fields (\ref{field1}) into field operators (such
that $b^{*}\to b^{\dagger}$) \cite{SW1}. These operators must fulfill the
standard commutation relations in the momentum representation among them the
non-vanishing ones are
\begin{equation}\label{com1}
[a({\bf p},\lambda), a^{\dagger}({\bf p}^{\,\prime},\lambda
')]=[b({\bf p},\lambda), b^{\dagger}({\bf p}^{\,\prime},\lambda ')]
=\delta_{\lambda\lambda '} \delta^3 ({\bf p}-{\bf p}^{\,\prime})\,.
\end{equation}
The field operators act on the Fock space supposed to have an unique vacuum
state $|0\rangle$ accomplishing
\begin{equation}
a({\bf p},\lambda)\,|\,0\rangle=0\,,\quad \langle
0|\,a^{\dagger}({\bf k},\lambda)=0\,,
\end{equation}
and similarly for $b$ and $b^{\dagger}$. The sectors with a given number of
particles have to be constructed using the standard methods, obtaining thus the
generalized momentum basis of the Fock space.

\subsection{Commutator functions}

The Green functions of the vector field are related to the partial commutator
functions (of positive or negative frequencies) defined as
\begin{equation}\label{comF}
D_{\mu\nu}^{(\pm)}(x,x')= i[A_{\mu}^{(\pm)}(x),A_{\nu}^{(\pm)\,\dagger}(x')]\,,
\end{equation}
and the total one, $D_{\mu\nu}=D^{(+)}_{\mu\nu}+D^{(-)}_{\mu\nu}$, which is a
real function since $[D_{\mu\nu}^{(\pm)}]^*=D_{\mu\nu}^{(\mp)}$. All these
functions are solutions of the field equations and obey the Lorentz condition
in both the sets of variables, $x$ and $x'$.  The properties of the commutator
functions can be deduced focusing only on the functions of positive
frequencies,
\begin{equation}
D^{(+)}_{\mu\nu}(x,x')= i\sum_{\lambda}\int d^3 p \,{\rm U}[{\bf
p},\lambda]_{\mu}(x){\rm U}[{\bf p},\lambda]_{\nu}(x')^*\,,
\end{equation}
derived from  Eqs. (\ref{field1}) and (\ref{tran}). According to Eqs.
(\ref{UU1}) and (\ref{UU2}) we obtain the mode integral expansions
\begin{eqnarray}
D^{(+)}_{ij}(x,x')&=&i\int d^3p\,\left[\left(\delta_{ij}-
\frac{p^ip^j}{p^2}\right) \alpha(t,p)\alpha(t',p)^*\right.\nonumber\\
&&\hspace*{28mm}\left.+\frac{p^ip^j}{p^2}\,\beta(t,p)\beta(t',p)^*\right]e^{i{\bf
p}({\bf x}-{\bf x}')}\,,\label{Dij}\\
D^{(+)}_{i0}(x,x')&=&i\int d^3p\,\frac{p^i}{p} \beta(t,p)\gamma(t',p)^*
e^{i{\bf p}({\bf x}-{\bf x}')}\,,\label{Di0}\\
D^{(+)}_{0i}(x,x')&=&i\int d^3p\,\frac{p^i}{p} \gamma(t,p)\beta(t',p)^*
e^{i{\bf p}({\bf x}-{\bf x}')}\,,\label{D0i}\\
D^{(+)}_{00}(x,x')&=&i\int d^3p\,\gamma(t,p)\gamma(t',p)^* e^{i{\bf p}({\bf
x}-{\bf x}')}\,,\label{D00}
\end{eqnarray}
which show that $D_{\mu\nu}^{(\pm)}(x,x')=D_{\mu\nu}^{(\pm)}(t,t',{\bf x}-{\bf
x}')$.

The above equations help us to deduce what happens at equal times, $t'=t$.
Indeed, bearing in mind that $D_{\mu\nu}$ are real functions and using the
identities given in Appendix A we find
\begin{eqnarray}
&&\left.D_{ij}(x,x')\right|_{t'=t}=0\,,\qquad
\left.D_{00}(x,x')\right|_{t'=t}=0\,,\label{first}\\
&&\left.D_{i0}(x,x')\right|_{t'=t}=\left.D_{0i}(x,x')\right|_{t'=t}=\frac{\omega^2
t^2}{m^2}\,\partial_i\,\delta^3({\bf x}-{\bf x}')\,,
\end{eqnarray}
and
\begin{eqnarray}
&&\left.\frac{1}{2}(\partial_t-\partial_{t'}) D_{ij}(x,x')\right|_{t'=t}=
\left(-\delta_{ij}+\frac{\omega^2
t^2}{m^2}\,\partial_i\partial_j\right)\,\delta^3({\bf x}-{\bf x}')\,,\\
&&\left.\frac{1}{2}(\partial_t-\partial_{t'}) D_{00}(x,x')\right|_{t'=t}=
\frac{\omega^2
t^2}{m^2}\,\Delta_{x}\, \delta^3({\bf x}-{\bf x}')\,,\\
&&\left.\frac{1}{2}(\partial_t-\partial_{t'})
D_{i0}(x,x')\right|_{t'=t}=\left.\frac{1}{2}(\partial_t-\partial_{t'})
D_{0i}(x,x')\right|_{t'=t}=0\,.\label{last}
\end{eqnarray}

The mode integrals (\ref{Dij})-(\ref{D00}) can be solved in terms of a scalar
function and some simple operators. This is just the commutator function of
positive frequencies of the scalar field {\em conformally}  coupled to the dS
gravity \cite{BD}, defined by the integral
\begin{equation}
{\cal D}^{(+)}(x,x')=\frac{i \pi\omega^2}{4}\,\frac{e^{-\pi
k}}{(2\pi)^3}\,(tt')^{3/2}  \int d^3 p\, e^{i({\bf x}-{\bf x}')\cdot{\bf
p}}H^{(1)}_{ik}(-pt)H^{(1)}_{ik}(-pt')^*
\end{equation}
which can be solved as \cite{PPP,PPG},
\begin{equation}
{\cal D}^{(+)}(x,x')=\frac{i m^2}{16 \pi}\,e^{-\pi k}{\rm sech}(\pi k)\,
_2F_1\left(\frac{3}{2}+ik,\frac{3}{2}-ik;2;1-\frac{y}{4}\right)\,,
\end{equation}
where the quantity
\begin{equation}
y(x,x')=\frac{-(t-t'-i\epsilon)^2+({\bf x}-{\bf x}')^2}{tt'}
\end{equation}
is related to the geodesic length between $x$ and $x'$. We note that the
function ${\cal D}^{(+)}$ satisfies the equation
\begin{equation}
\left(\partial_t^2-\frac{2}{t}\,\partial_t-\Delta_x+\frac{\mu^2+2}{t^2}\right){\cal
D}^{(+)}(x,x')=0
\end{equation}
(and similarly for $x'$). With its help and using Eqs. (\ref{alpha}),
(\ref{gamma}) and (\ref{eqc}) we can write:
\begin{eqnarray}
D^{(+)}_{ij}(t,t',{\bf
x})&=&\frac{1}{\omega^2tt'}\left(\delta_{ij}-\frac{\partial_i\partial_j}
{\Delta}\right){\cal D}^{(+)}(t,t',{\bf x})\nonumber\\
&&+\frac{1}{m^2}\,\frac{\partial_i\partial_j}{\Delta}\left(\partial_t-\frac{2}{t}\right)
\left(\partial_{t'}-\frac{2}{t'}\right){\cal D}^{(+)}(t,t',{\bf x})\,,\label{DDij}\\
D^{(+)}_{i0}(t,t',{\bf
x})&=&-\frac{1}{m^2}\,\partial_i\left(\partial_t-\frac{2}{t}\right)
{\cal D}^{(+)}(t,t',{\bf x})\\
D^{(+)}_{0j}(t,t',{\bf
x})&=&\frac{1}{m^2}\,\partial_i\left(\partial_{t'}-\frac{2}{t'}\right)
{\cal D}^{(+)}(t,t',{\bf x})\\
D^{(+)}_{00}(t,t',{\bf x})&=&-\frac{1}{m^2}\,\Delta {\cal D}^{(+)}(t,t',{\bf
x})\,.
\end{eqnarray}
Similar formulas can be derived for $D_{\mu\nu}^{(-)}$ and $D_{\mu\nu}$ which
are related to the scalar functions ${\cal D}^{(-)}=[{\cal D}^{(+)}]^*$ and
${\cal D}={\cal D}^{(+)}+{\cal D}^{(-)}$ respectively.

We succeeded thus to express all the commutator functions of the vector field
in terms of some operators acting on scalar functions of $y$. A similar result
was obtained recently for the Dirac field whose anti-commutator functions
resulted from the canonical quantization are given by differential
matrix-operators acting on specific scalar functions depending on the geodesic
length \cite{KoPro}. Moreover, in both these cases there are factors depending
on powers of $t$ and $t'$ which show that the entire commutator or
anti-commutator functions are no longer genuine functions of $y$ (or of the
geodesic length) and its derivatives. Notice that, in addition, the
non-differential operator $\Delta^{-1}$ of Eq. (\ref{DDij}) affects the
dependence on spaces variables too.

\subsection{Green functions}

According to the standard procedure, we now define the retarded ($R$), advanced
($A$), and (causal) Feynman ($F$) propagators,
\begin{eqnarray}
\tilde D^R_{\mu\nu}(x,x')&=&\theta(t-t')D_{\mu\nu}(x,x')\label{DR}\,, \\
\tilde D^A_{\mu\nu}(x,x')&=&-\theta(t'-t)D_{\mu\nu}(x,x')\label{DA}\,,\\
\tilde D^F_{\mu\nu}(x,x')&=& i\langle 0|\,T\,[A_{\mu}(x) A_{\nu}^{\dagger}(x')]
\,|\,0\rangle\nonumber\\
&=& \theta (t-t') D_{\mu\nu}^{(+)}(x,x')-\theta(t'-t)D_{\mu\nu}^{(-)}(x,
x')\,.\label{DF}
\end{eqnarray}
The corresponding Green functions,
\begin{equation}\label{GRAF}
G^{R/A/F}_{\mu\nu}(x.x')= \frac{\omega^2
t^2}{m^2}\,\delta_{\mu}^0\delta^0_{\nu}\,\delta^4(x-x')+\tilde
D_{\mu\nu}^{R/A/F}(x,x')\,,
\end{equation}
satisfy the equation
\begin{equation}\label{EQG}
\eta^{\alpha\beta}\partial_{\alpha}\left[\partial_{\beta}G_{\mu\nu}(x,x')
-\partial_{\mu}G_{\beta\nu}(x,x')\right]+\frac{m^2}{\omega^2 t^2}\,
G_{\mu\nu}(x,x')=\eta_{\mu\nu}\delta^4(x-x')\,.
\end{equation}
This can be proved by using the identities
$\partial_t^2[\theta(t)f(t)]=\delta(t)\partial_t f(t)-\delta(t) f(t)\partial_t$
and $\partial_t[\delta(t)f(t)]= - \delta(t)f(t)\partial_t$, the artifice
$\partial_t f(t-t')=\frac{1}{2}(\partial_t-\partial_{t'}) f(t-t')$ and Eqs.
(\ref{first})-(\ref{last}). In addition,  the Lorentz condition yields
\begin{equation}\label{LLL}
\eta^{\alpha\beta}\partial_{\alpha}\left[\frac{1}{\omega^2
t^2}\,G_{\beta\mu}(x,x') \right]=\frac{1}{m^2}\,\partial_{\mu}\delta^4(x-x')\,.
\end{equation}
In the flat limit these propagators and Green functions become the well-known
ones of special relativity  \cite{BS} as we briefly present in Appendix B.

Different Green functions can be defined by changing the gauge. The transverse
ones, obeying the exact Lorentz condition,
\begin{equation}\label{lori}
\partial_{\mu}\left[\sqrt{g(x)}g^{\mu\nu}(x)G^{tr}_{\nu\sigma}(x,x')\right]=
\partial_{\mu}'\left[\sqrt{g(x')}g^{\mu\nu}(x')G^{tr}_{\sigma\nu}(x,x')\right]=0 \,,
\end{equation}
can be defined as
\begin{equation}\label{Gtr}
G^{tr}_{\mu\nu}(x,x')=G_{\mu\nu}(x,x')+\frac{1}{m^2}
\partial_{\mu}\partial_{\nu}' G_0(x,x')\,,
\end{equation}\label{G0}
where $G_0$ is the massless scalar Green function which satisfies
\begin{equation}
\partial_{\mu}\left[\sqrt{g(x)} g^{\mu\nu}(x)\partial_{\nu}G_0(x,x')\right]=\delta^4(x-x')\,.
\end{equation}
According to Eqs. (\ref{EQG}) and (\ref{G0}) we find that the equation of the
transverse Green functions,
\begin{eqnarray}
&&\sqrt{g(x)}g^{\alpha\beta}(x)\partial_{\alpha}\left[\partial_{\beta}G^{tr}_{\mu\nu}(x,x')
-\partial_{\mu}G^{tr}_{\beta\nu}(x,x')\right]+m^2\,
\sqrt{g(x)}\,G_{\mu\nu}^{tr}(x,x')\nonumber\\
&&\hspace*{20mm}=g_{\mu\nu}(x)\delta^4(x-x')+\sqrt{g(x)}\,
\partial_{\mu}\partial_{\nu}'G_0(x,x')\,,\label{EQGtr}
\end{eqnarray}
coincides to that of Ref. \cite{PPV1} where one uses the same Lorentz condition
(\ref{lori}). In the flat limit this equation takes the usual form
(\ref{EQG11}).

Finally, we must stress that the transverse Green functions obtained here are
different from those derived in \cite{PPV1} even though all of them  satisfy
the same equation (\ref{EQGtr}). This is because the canonical quantization we
use generates  propagators that do not depend only on $y$ and its derivatives
as those of Ref. \cite{PPV1} which are forced to do this by definition.

\section{One-particle operators}

The canonical quantization procedure allows us to construct the one-particle
operators associated to the symmetry generators staring directly with the
conserved quantities (\ref{CX}). More precisely,  for each  generator $X$ we
assume that there exist a corresponding one-particle operator defined as
\begin{equation}\label{opo}
{\cal X}=:\langle A|\, X A\rangle:
\end{equation}
respecting the normal ordering of the operator products \cite{SW1}. The obvious
algebraic properties
\begin{equation}\label{algXX}
[{\cal X}, A_{\mu}(x))]=-(X A)_{\mu}(x)\,, \quad [{\cal X}, {\cal
Y}\,]=:\langle A |\, [X,Y]A\,\rangle:
\end{equation}
result from the quantization method adopted here.

However, there are many other conserved operators which do not have
corresponding differential operators at the level of quantum mechanics.  The
simplest examples are the operators of number of particles and antiparticles
respectively,
\begin{equation}
{\cal N}_{pa}=\int d^3p\, \sum_{\lambda} a^{\dagger}({\bf p},\lambda) a({\bf
p},\lambda)\,,\quad {\cal N}_{ap}=\int d^3p\,\sum_{\lambda} b^{\dagger}({\bf
p},\lambda) b({\bf p},\lambda)\,,
\end{equation}
which give the charge operator ${\cal Q}=:\langle A|\,A\rangle:={\cal
N}_{pa}-{\cal N}_{ap}$ and the operator of total number of particles, ${\cal
N}={\cal N}_{pa}+{\cal N}_{ap}$.

The principal conserved one-particle operators are ${\cal Q}$, the components
of momentum operator,
\begin{equation}
{\cal P}^i=:\langle A|\,  P^i A\rangle:=\int d^3p\, p^i \sum_{\lambda}[
a^{\dagger}({\bf p},\lambda) a({\bf p},\lambda)+b^{\dagger}({\bf p},\lambda)
b({\bf p},\lambda)]\,,
\end{equation}
and the Pauli-Lubanski operator,
\begin{equation}
{\cal W}=:\langle A |\, W A\rangle:\,=\int d^3p\,p
 \sum_{\lambda}\lambda\,[
a^{\dagger}({\bf p},\lambda) a({\bf p},\lambda)+b^{\dagger}({\bf p},\lambda)
b({\bf p},\lambda)]\,.
\end{equation}
The  complete set of commuting operators $\{{\cal Q},{\cal P}^i,{\cal W}\}$
determines the momentum basis of the Fock space.

The other one-particle operators are not diagonal in this basis but
can be written in closed forms.  For example, we can write the
Hamiltonian operator in the momentum basis starting with the
identity
\begin{equation}
\left(H{\rm U}[{\bf p},\lambda]\right)_{\mu}(x)=-i\omega
\left(p^i\partial_{p_i}+{\frac{3}{2}}\right){\rm U}[{\bf
p},\lambda]_{\mu}(x)\,.
\end{equation}
The final result,
\begin{equation}\label{Hpp}
{\cal H}=\frac{i\omega}{2}\int d^3p\, p^i  \sum_{\lambda}\,[a^{\dagger}({\bf
p},\lambda)\stackrel{\leftrightarrow}{\partial}_{p_i} a({\bf p},\lambda) +
b^{\dagger}({\bf p},\lambda)\stackrel{\leftrightarrow}{\partial}_{p_i} b({\bf
p},\lambda)]\,,
\end{equation}
is similar with those obtained for the scalar \cite{CS}, Maxwell \cite{Max} and
Dirac \cite{CD1} fields on $(M,g)$.

Another example is the total angular momentum whose components can be easily
represented in the momentum basis. Indeed, according to Eqs. (\ref{XA}) and
(\ref{J}), we obtain the closed form
\begin{eqnarray}
{\cal J}_l&=&-\frac{i}{2}\,\varepsilon_{lij}\int d^3p\,\left\{ p^i
\sum_{\lambda}\, [a^{\dagger}({\bf
p},\lambda)\stackrel{\leftrightarrow}{\partial}_{k_j} a({\bf
p},\lambda)+b^{\dagger}({\bf
p},\lambda)\stackrel{\leftrightarrow}{\partial}_{k_j} b({\bf
p},\lambda)]\right.\nonumber\\
&&+\left.\sum_{\lambda \lambda'}\, \vartheta^{ij}_{\lambda \lambda'}({\bf
p})\,[a^{\dagger}({\bf p},\lambda) a({\bf p},\lambda')+b^{\dagger}({\bf
p},\lambda) b({\bf p},\lambda')]\right\}\,,
\end{eqnarray}
where we denote
\begin{equation}
\vartheta^{ij}_{\lambda \lambda'}({\bf p})=2 e_i({\bf p},\lambda)^*e_j({\bf
p},\lambda')+ p^i\sum_l e_l({\bf
p},\lambda)^*\stackrel{\leftrightarrow}{\partial}_{p_j} e_l({\bf
p},\lambda')\,.
\end{equation}
Notice that Eq. (\ref{eee}) helps us to verify the identity ${\cal
W}=\sum_i{\cal P}_i{\cal J}_i$.

\section{Conclusions}

In this paper we succeeded to built the quantum theory of the massive charged
vector field minimally coupled to the gravity of the dS expanding universe. Our
approach is based on a complete set of commuting operators which determines in
turn the fundamental solutions of given momentum and polarization. These form a
complete set of orthonormalized solutions with respect to the relativistic
scalar product. Under such circumstances, the method of canonical quantization
was used for constructing the Fock space and the principal one-particle
operators.

The one-particle operators we derived here have similar structures and
properties as in the scalar \cite{CS}, Dirac \cite{CD1} or Maxwell \cite{Max}
theories. It is remarkable that the expansion in the momentum basis of the
Hamiltonian operator (which is not diagonal in this basis) can be done as in
the above mentioned cases \cite{CS,CD1,Max}, using similar formulas. This
indicates that our definition of the Hamiltonian operator \cite{CD1} is correct
despite of some doubts on its existence appeared in literature \cite{EW}. We
specify that this operator is globally defined by the Killing vector $K_{05}$,
but it makes sense only inside the light-cones where it is always time-like. In
other words, the energy is well-defined  wherever an observer can do physical
measurements.

It is worth pointing out that our approach has good massless an flat limits. In
the massless limit this leads to the quantum theory of the free Maxwell field
in Coulomb gauge on the dS expanding universe we proposed recently \cite{Max}.
The flat limit of our theory recovers all the well-know results of the Proca
theory on Minkowski backgrounds. For this reason,  our Green functions
(\ref{GRAF}) have the same structure (\ref{GRAF1}) and similar properties to
those of the corresponding Green functions on the flat spacetime. However,
despite of these similarities there are major differences. Apart from the
analytical expressions, the principal difference is that on dS spacetimes the
vector Green functions can not be related to the scalar propagators such as
done in Eq. (\ref{GM})and, consequently, there are no propagator equations
similar to Eq. (\ref{GM1}).

In other respects, it is obvious that our transverse Green functions
(\ref{Gtr}) differ from the maximally symmetric two-point functions proposed by
Allen and Jacobson \cite{PPV}  but satisfy the same equation and Lorentz
condition as the propagators constructed axiomatically by Tsamis and Woodard
\cite{PPV1}. The difference is that our propagators are no longer functions
only of $y$ and its derivatives while those of Ref. \cite{PPV1} have this
property. This suggests that our transverse Green functions we obtained using
the canonical quantization are new solutions of the transverse equation
(\ref{EQGtr}).

The Proca theory on the dS expanding universe we presented here completes our
previous works \cite{CS,CD1,CD2,Max} opening thus the perspective to a
realistic quantum theory on the dS expanding universe involving scalar, vector
and spinor fields.  This theory must be based on trustworthy results given by
the canonical quantization and perturbation theory in the reduction formalism.
In this framework one could deal with more sophisticated methods but now
preserving the minimum requirements of rigor and consistency.

\subsection*{Acknowledgments}

We are grateful to Erhardt Papp  for interesting and useful discussions on
closely related subjects.

\appendix

\subsection*{Appendix A: Properties of some Bessel functions}

Let us consider the Hankel functions $H^{(1,2)}_{\nu}(s)$ in the
special case when $\nu=i k$ and denote
\begin{equation}
Z(s)=e^{-\pi k/2}H^{(1)}_{ik}(s)\,,\quad Z^*(s)=e^{\pi
k/2}H^{(2)}_{ik}(s)\,.
\end{equation}
Then,  by using the Wronskian $W$ of the Bessel functions \cite{AS} we find
that
\begin{equation}\label{ZuZu}
Z^*(s)
\stackrel{\leftrightarrow}{\partial_{s}}Z(s)=W[H_{ik}^{(2)},H_{ik}^{(1)}]=\frac{4i}{\pi
s}\,.
\end{equation}
Starting with this property and Eq. (\ref{Lor}) we find the useful identities
\begin{eqnarray}
\alpha(t,p)^*\stackrel{\leftrightarrow}{\partial_{t}}\alpha(t,p)&=&\frac{4i}{\pi}|N_1|^2
=\frac{i}{(2\pi)^3}\,,\\
\gamma(t,p)^*\stackrel{\leftrightarrow}{\partial_{t}}\gamma(t,p)&=&\frac{4i}{\pi}
t^2|N_2|^2 = \frac{i}{(2\pi)^3} \frac{\omega^2 t^2}{m^2}\,p^2\,,\\
\beta(t,p)^*\stackrel{\leftrightarrow}{\partial_{t}}\beta(t,p)&=&\left(1+\frac{\mu^2}{p^2
t^2}\right)\gamma(t,p)^*\stackrel{\leftrightarrow}{\partial_{t}}\gamma(t,p)\nonumber\\
&=&\frac{4i}{\pi}\left(t^2+\frac{\mu^2}{p^2}\right)|N_2|^2 =\frac{i}{(2\pi)^3}
\left(1+\frac{\omega^2 t^2}{m^2}\,p^2\right)\,,
\end{eqnarray}
which helped us to find the normalization factors (\ref{N1N2})
according to Eqs. (\ref{orto}) and (\ref{AA}). In addition, from Eq.
(\ref{Lor}) we deduce
\begin{equation}
\Re \left[\gamma(t,p)^*\beta(t,p)\right]=\frac{1}{(2\pi)^3}\frac{\omega^2
t^2}{m^2} \frac{p}{2}\,.
\end{equation}

\subsection*{Appendix B: The flat limit}

In the flat limit, when $\omega\to 0$ and $\omega t\to -1$, Eq. (\ref{EQG})
becomes the standard equation of the vector Green functions of special
relativity,
\begin{equation}\label{EQG1}
\eta^{\alpha\beta}\partial_{\alpha}\left[\partial_{\beta}G_{\mu\nu}(x)
-\partial_{\mu}G_{\beta\nu}(x)\right]+m^2
G_{\mu\nu}(x)=\eta_{\mu\nu}\delta^4(x)\,,
\end{equation}
whose solutions \cite{BS},
\begin{equation}\label{GM}
G_{\mu\nu}(x)=\left(\eta_{\mu\nu}+\frac{1}{m^2}\partial_{\mu}\,\partial_{\nu}\right)\tilde
D(x)
\end{equation}
depend on the scalar propagator $\tilde D$ which obeys $(\partial^2+m^2)\tilde
D(x)=\delta^4(x)$. These Green functions can be written as \cite{BS}
\begin{equation}\label{GRAF1}
G_{\mu\nu}(x)= \frac{1}{m^2}\,\delta_{\mu}^0\delta^0_{\nu}\,\delta^4(x)+\tilde
D_{\mu\nu}(x)\,,
\end{equation}
where the  vector propagators $\tilde D_{\mu\nu}$ are defined as in Eqs.
(\ref{DR})-(\ref{DF}). According to Eq. (\ref{GM}), one can replace Eq.
(\ref{EQG1}) by the well-known one,
\begin{equation}\label{GM1}
(\partial^2+m^2)G_{\mu\nu}(x)=\left(\eta_{\mu\nu}
+\frac{1}{m^2}\partial_{\mu}\,\partial_{\nu}\right)\delta^4(x)\,,
\end{equation}
and  show that the Lorentz condition yields,
\begin{equation}\label{trans}
\partial^{\mu} G_{\mu\nu}(x) = \frac{1}{m^2}\,\partial_{\nu}\delta^4
(x)\,.
\end{equation}

The transverse propagator,
\begin{equation}\label{Gtr1}
G^{tr}_{\mu\nu}(x)=G_{\mu\nu}(x)-\frac{1}{m^2}
\partial_{\mu}\partial_{\nu} \tilde D_0(x)\,,
\end{equation}
depend on the massless scalar propagator $\tilde D_0$. It satisfies the exact
Lorentz condition, $\partial^{\mu} G_{\mu\nu}^{tr}(x) = 0$, and the equation
\begin{equation}\label{EQG11}
\eta^{\alpha\beta}\partial_{\alpha}\left[\partial_{\beta}G_{\mu\nu}^{tr}(x)
-\partial_{\mu}G_{\beta\nu}^{tr}(x)\right]+m^2
G_{\mu\nu}^{tr}(x)=\eta_{\mu\nu}\delta^4(x)-\partial_{\mu}\partial_{\nu} \tilde
D_0(x)\,.
\end{equation}


\begin{thebibliography}{20}

\bibitem{BD}
N. D. Birrel and P. C. W. Davies,  {\em Quantum Fields in Curved Space}
(Cambridge University Press, Cambridge 1982).

\bibitem{WALD}
R. M. Wald,  {\it General Relativity}, (Univ. of Chicago Press, Chicago and
London, 1984)

\bibitem{V}
A. Higuchi, {\em Class. Quant. Gravity} {\bf 4}, 712 (1987); D. Bini, G.
Esposito and R. V. Montaquila, {\tt arXiv:0812.1973}.


\bibitem{PPP}
N. A. Chernikov and E. A. Tagirov, {\em Ann. Inst H. Poincar\' e} {\bf IX} 1147
(1968)

\bibitem{PPG}
P. Candelas and D. J. Raine, {\em Phys. Rev. D} {\bf 12} 965 (1975); J. S.
Dowker and J. S. Critchely, {\em Phys. Rev. D} {\bf 13}, 224 (1976); T. S.
Bunch and P. C. W. Davies, {\em Proc. R. Soc. Lond. A} {\bf 360}, 117 (1978);

\bibitem{PPV}
B. Allen and T. Jacobson {\em Commun. Math. Phys.} {\bf 103}, 669 (1986).

\bibitem{PPV1}
N. C. Tsamis and R. P. Woodard, {\em J.Math.Phys.} {\bf 48},  052306 (2007),
{\tt gr-qc/0608069}.

\bibitem{Dav}
O. Bertolami and D. F. Mota {\em Phys. Lett. B} {\bf 455}, 96 (1999), {\tt
gr-qc/9811087}

\bibitem{Woo}
T. Prokopek, O. T\" ornkvist and R. P. Woodard, {\em Phys Rev. Lett.} {\bf 89},
101301 (2002);   T. Prokopec, N. C. Tsamis and R. P. Woodard, {\em Class.
Quant. Grav.} {\bf 24} (2007), {\tt gr-qc/0607094}; ibid, {\em Phys. Rev. D}
{\bf 79}, 043523 (2008).

\bibitem{CML}
B. Carter and R. G. McLenaghan, {\em Phys. Rev. D} {\bf 19}, 1093 (1979).

\bibitem{ES}
I. I. Cot\u aescu, {\em J. Phys. A: Math. Gen.} {\bf 33}, 9177 (2000).

\bibitem{CSP}
I. I. Cot\u aescu, {\em Mod. Phys. Lett. A} {\bf 22}, 2965 (2007).

\bibitem{CS}
I. I. Cot\u aescu, C. Crucean and A. Pop, {\em Int. J. Mod. Phys A}
{\bf 23}, 2563 (2008), {\tt arXiv:0802.1972 [gr-qc]}

\bibitem{CD1}
I. I. Cot\u aescu, {\em Phys. Rev. D} {\bf 65}, 084008 (2002).

\bibitem{CD2}
I. I. Cot\u aescu and C. Crucean, {\em Int. J. Mod. Phys A} {\bf
23}, 3703 (2008); I. I. Cot\u aescu, {\tt arXiv:0711.0816 [gr-qc]}.

\bibitem{Max}
I. I. Cot\u aescu and C. Crucean {\tt arXiv:0806.2515 [gr-qc]}.

\bibitem{BS}
N. N. Bogoliubov and D. V. Shirkov, {\it Introduction to the theory of
quantized fields} (Interscience Publ. N. Y., 1959).

\bibitem{SW1}
S. Weinberg, {\it The Quantum Theory of Fields}  (Univ. Press, Cambridge,
1995).

\bibitem{KoPro}
J. F. Koksma and T. Protopek, {\em Class. Quantum Grav.} {\bf 26}, 125003
(2009).

\bibitem{EW}
E. Witten, {\tt hep-th/0106109}

\bibitem{AS}
M. Abramowitz and I. A. Stegun, {\it Handbook of Mathematical Functions}
(Dover, 1964)



\end{thebibliography}
\end{document}